\documentclass[prb, aps,amsmath,amssymb,amsfonts, twocolumn, superscriptaddress]{revtex4}
\usepackage[german,american,english]{babel}
\usepackage{graphicx}% Include figure files
\usepackage{graphics}% Include figure files
\usepackage{dcolumn}% Align table columns on decimal point
\usepackage{bm}% bold math
\usepackage{amssymb}
\usepackage{amsmath}
\usepackage{amsfonts}
\usepackage{epsfig}

\newcommand {\ket} [1] {| #1 \rangle}

\newcommand {\tbkt} [3] {\langle #1 | #2 | #3 \rangle}

 \newcommand {\beq}{\begin{equation}}
\newcommand {\eeq}{\end{equation}}
\begin{document}
\title{Coulomb interaction and valley-orbit coupling in Si quantum dots}
\author{Luyao Jiang}
\affiliation{ICQD, Hefei National Laboratory for Physical Sciences at the Microscale, University of Science and Technology of China, Hefei 230026, Anhui, China}

\author{C. H. Yang}
\affiliation{ARC Centre of Excellence for Quantum Computation and Communication Technology, School of Electrical Engineering \& Telecommunications, The University of New South Wales, Sydney 2052, Australia}

\author{Zhaodi Pan}
\affiliation{ICQD, Hefei National Laboratory for Physical Sciences at the Microscale, University of Science and Technology of China, Hefei 230026, Anhui, China}

\author{Alessandro Rossi}
\affiliation{ARC Centre of Excellence for Quantum Computation and Communication Technology, School of Electrical Engineering \& Telecommunications, The University of New South Wales, Sydney 2052, Australia}

\author{Andrew S. Dzurak}
\affiliation{ARC Centre of Excellence for Quantum Computation and Communication Technology, School of Electrical Engineering \& Telecommunications, The University of New South Wales, Sydney 2052, Australia}

\author{Dimitrie Culcer}
\affiliation{School of Physics, The University of New South Wales, Sydney 2052, Australia}
\affiliation{ICQD, Hefei National Laboratory for Physical Sciences at the Microscale, University of Science and Technology of China, Hefei 230026, Anhui, China}
\begin{abstract}
The valley-orbit coupling in a few-electron Si quantum dot is expected to be a function of its occupation number $N$. We study the spectrum of multivalley Si quantum dots for $2 \le N \le 4$, showing that, counterintuitively, electron-electron interaction effects on the valley-orbit coupling are negligible. For $N=2$ they are suppressed by valley interference, for $N=3$ they vanish due to spinor overlaps, and for $N = 4$ they cancel between different pairs of electrons. To corroborate our theoretical findings, we examine the experimental energy spectrum of a few-electron metal-oxide-semiconductor quantum dot. The measured spin-valley state filling sequence in a magnetic field reveals that the valley-orbit coupling is definitively unaffected by the occupation number.
\end{abstract}

\maketitle

\section{Introduction}

The requirements of quantum bit (qubit) scalability and long coherence times have brought solid-state spin systems to the fore of quantum computation \cite{Kane_Nature98, Loss_PRA98, Petta_Science05, Koppens_PRL08, Awsch_Qbt_Rvw_Sci13}. Realizing spin qubits in Si quantum dots is particularly promising thanks to the remarkable coherence properties of Si \cite{Abe_PRB04, Tyryshkin_JPC06, Morton_Si_QC_QmLim_Nat11, Prada_PRB08, Wilamowski_Si/SiGeQW_Rashba_PRB02, Witzel_AHF_PRB07}, and significant experimental progress has lately been reported \cite{Zwanenburg_SiQmEl_RMP13, Hanson_RMP07, Taylor_PRB07, Ono_Science02}. The valley degree of freedom of Si \cite{Ando_PRB79} is vital in quantum computing \cite{Friesen_PRB04, Hada_JJAP04, Friesen_PRB10, Simmons_DQD_SpinBloc_LET_PRB10, Xiao_MOS_SpinRelax_PRL10, Culcer_ValleyQubit_PRL12, Palyi_PRL11, Culcer_PRB10, Goswami_NP07, Lai_PRB06, Takashina_PRL06, Lim_SiQD_SpinFill_NT11, Yang_SpinVal_NC13, Nestoklon_PRB06, Boykin_APL04, Srinivasan_APL08, Saraiva_EMA_PRB11, SDS_SiQD_Hubbard_PRB11, Gamble_VlyRlx_PRB12}: it enables valley-based information processing \cite{Culcer_ValleyQubit_PRL12, Wu_Culcer_PRB2012} and resonance \cite{Palyi_PRL11}, but hampers spin qubits \cite{Culcer_PRB10}. A valley-orbit coupling (VOC) causes valley states to hybridize into \textit{valley eigenstates} \cite{Goswami_NP07, Lai_PRB06, Takashina_PRL06, Lim_SiQD_SpinFill_NT11, Yang_SpinVal_NC13, Nestoklon_PRB06, Boykin_APL04, Srinivasan_APL08, Saraiva_EMA_PRB11}, and the biggest issues at present are an understanding of the magnitude of the VOC, its response to applied fields, and its sensitivity to interactions with the environment and among electrons.

The VOC of singly occupied quantum dots stems from the interface potential \cite{Saraiva_EMA_PRB11}, yet for few-electron quantum dots, interactions contribute to the VOC. Since Coulomb repulsion depends on the spatial separation of electrons, it is often appreciable in quantum dots, where electrons are strongly confined. The on-site interaction $u$ can exceed the confinement energy, and in Si double quantum dots repulsion contributes sizably to interdot tunneling \cite{Culcer_PRB10}. The relative contribution of interactions to the VOC of few-electron quantum dots is an issue of substantial conceptual and practical significance. 

% We need thorough justification to regard it as one-particle. Interactions are important in the VOC in the QHE regime leading to valley gaps analogous to spin gaps.

This paper is a theoretical and experimental study of interaction effects on the VOC in a single Si quantum dot with $1 \le N \le 4$. For $N=2$, we show that the Coulomb interaction effect on the VOC is suppressed by valley interference, while for $N= 3$ and $N=4$ interaction terms vanish due to spinor overlaps. Experimental data supports this finding, showing no evidence of interactions in the valley-orbital spectrum of Si quantum dots with $1 \le N \le 3$. We argue that these observations apply beyond $N = 4$.

The outline of this paper is as follows. In Sec.\ \ref{sec:model} we introduce the model of the quantum dot, and discuss the spectrum for $1 \le N \le 4$. In Sec.\ \ref{sec:expt} we compare our theoretical findings with experimental data. Sec.\ \ref{sec:disc} is devoted to a discussion of the underlying physics, and is followed by a summary and conclusions.

\section{Model}
\label{sec:model}

\begin{figure}[tbp]
\centering
\includegraphics[width=0.8\columnwidth]{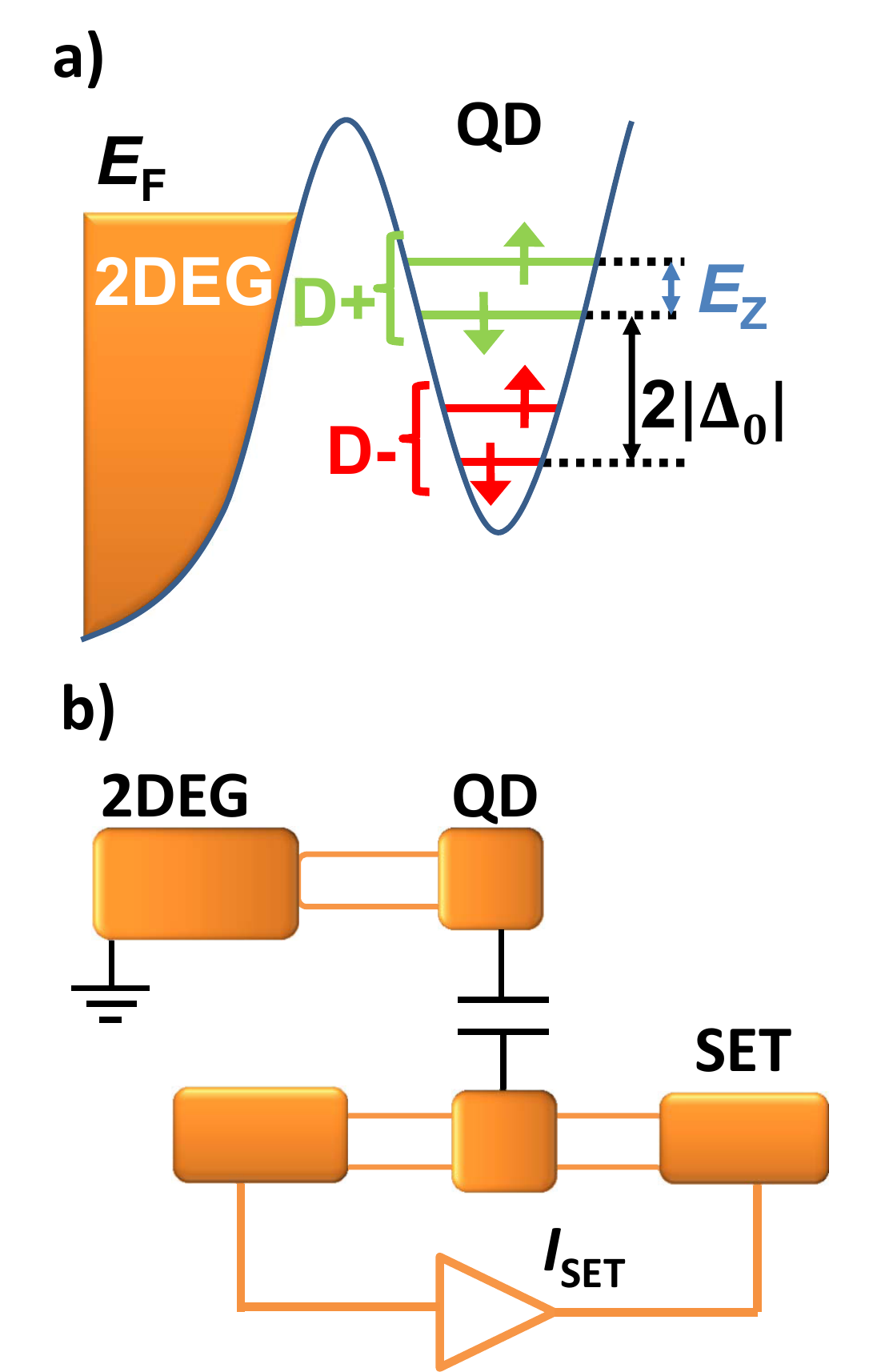}
\caption{(a) Energy diagram of the four lower spin-valley single-particle levels in the quantum dot. The Zeeman splitting is assumed to be smaller than the valley-orbit coupling. A tunnel barrier separates the dot from a 2-dimensional electron gas (2DEG), acting as a reservoir. (b) Device architecture used for the experiments. A single-lead quantum dot (top) is capacitively coupled to a single-electron transistor charge sensor (bottom). Regions filled in orange represent electron layers, non-filled regions are tunnel barriers. The readout current signal $I_{SET}$ is sensitive to the dot's charge state via capacitive effects.}
\label{fig:explain}
\end{figure}

We consider a dot located at the origin with Fock-Darwin radius $a$. The confinement potential
\begin{equation}
V_D(x,y,z)=\frac {\hbar} {2m^{\ast}a^2}(\frac {x^2+y^2} {a^2})+U_0\theta(-z)+eFz,
\end{equation}
with the electron charge $-e$, in-plane effective mass $m^\ast$, interface potential $V_z=U_0\theta(-z)$, $\theta$ the Heaviside function, and gate electric field $F$. The electron wave functions $D_{\xi}(x,y,z)=\phi_D(x,y)\psi(z)u_{\xi}(\bm r)e^{ik_{\xi}z}$, the valley index $\xi=\{z,-z\}$, and $k_{z,- z}=\pm k_0=\pm0.85(2\pi/a_{Si})$, with $a_{Si} = 5.43 \AA$ the Si lattice constant. The envelopes $\phi_D(x,y)=\frac 1 {a \sqrt{\pi}}e^{-(x^2+y^2)/2a^2}$, while $\psi(z) = Nz_0e^{k_bz/2}\theta(-z)+N(z+z_0)e^{-k_{Si}z/2}\theta(z)$ is a variational envelope function, with $k_{Si}$ a variational parameter, $k_b$ fixed, and $N$ the normalization \cite{Culcer_ValleyQubit_PRL12}. Given the twofold spin and valley degrees of freedom, the lowest single-particle levels can accommodate up to 4 electrons [see Fig.~\ref{fig:explain} (a)]. For $N \le 4$ the number of many-electron states is the combination $C^4_{N}$. The case $N = 1$ has been studied at length, and the \textit{bare} (single-particle) VOC is $\Delta_0 \equiv |\Delta_0| \, e^{-i\phi_0} = \tbkt{D_z}{U_0 \, \theta(- z) + e F z}{D_{-{z}}}$. The valley-orbit \textit{splitting} is 2$|\Delta_0|$, and the valley \textit{eigenstates} are $D_\pm = (1/\sqrt{2}) \, (D_z \pm e^{i\phi_0} D_{- z})$ \cite{Culcer_ValleyQubit_PRL12}.

\subsection{Two-electron case}

For ${N=2}$ the Hamiltonian $H_{2e}=T^{(1)}+V_D^{(1)}+T^{(2)}+V_D^{(2)}+V_{ee}$ where superscripts $i \in \{1, 2\}$ label electrons at ${\bm r}_1$ and ${\bm r}_2$, $T$ is the kinetic energy, and $V_{ee}=\frac {e^2} {4\pi\epsilon_0 \epsilon_r |\bm {r_1}-\bm {r_2}|}$ is the Coulomb interaction, with $\epsilon_0$ the permittivity and $\epsilon_r$ the relative permittivity. The $C^4_2 = 6$ two-particle states are 3 spin singlets ($\phi^S$) and 3 spin triplets ($\phi^T$),
\begin{flalign}\label{defs}
\begin{split}
&\phi^S_{\pm z, \pm z} =  D_{\pm z}^{(1)}D_{\pm z}^{(2)}\chi_S\\
&\phi^{S, T}_{mix} =  \frac{1}{\sqrt 2} \, (D_z^{(1)}D_{- z}^{(2)} \pm D_z^{(2)}D_{-z}^{(1)} ) \, \chi_{S, T},
\end{split}
\end{flalign}
with $\chi_S$ and $\chi_T$ \textit{spin} the singlet and triplet spin wave functions, given by 
\begin{equation}
\arraycolsep 0.3 ex
\begin{array}{rl}
\displaystyle \ket{\chi_S} = & \displaystyle \frac{1}{\sqrt{2}} \, ( \ket{\uparrow^{(1)}\downarrow^{(2)}} - \ket{\downarrow^{(1)}\uparrow^{(2)}}) \\ [3ex]
\displaystyle \ket{\chi_{T, \uparrow \uparrow}} = & \displaystyle \ket{\uparrow^{(1)}\uparrow^{(2)}} \\ [3ex]
\displaystyle \ket{\chi_{T, \downarrow \downarrow}} = & \displaystyle \ket{\downarrow^{(1)}\downarrow^{(2)}} \\ [3ex]
\displaystyle \ket{\chi_{T, 0}} = & \displaystyle \frac{1}{\sqrt{2}} \, ( \ket{\uparrow^{(1)}\downarrow^{(2)}} + \ket{\downarrow^{(1)} \uparrow^{(2)}}). 
\end{array}
\end{equation}
We use $\ket{\chi_T}$ generically for any of the three triplet wave functions.  

\begin{figure}[tbp]
\centering
\includegraphics[width=\columnwidth]{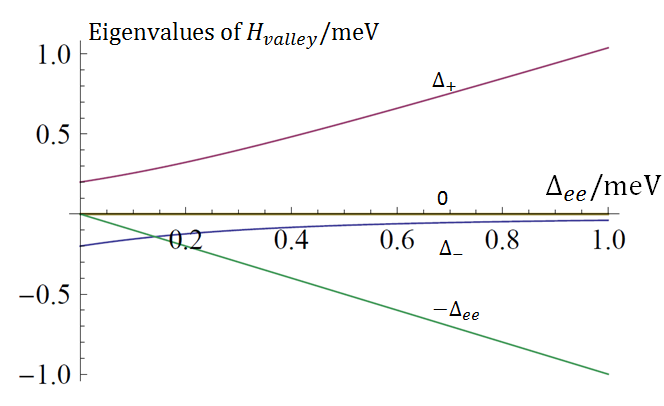}
\caption{Eigenvalues of $H_{valley}$ as functions of $\Delta_{ee}$ for a quantum dot with $N=2$ electrons, with $\Delta_0 = 0.05$ meV. At large $\Delta_{ee}$ the ground state is a spin triplet (green line).}
\label{fig:control}
\end{figure}

In the basis $\{\phi^S_{zz}, \phi^S_{-z - z}, \phi^S_{mix} ,\phi^T_{mix}\}$, the two-electron Hamiltonian can be written as $H_{2e}=2(\varepsilon_0+u)\openone+H_{val}$, with $\varepsilon_0=\langle D_{\xi_1}^{(i)}|[T^{(i)} + V_D]|D_{\xi_1}^{(i)}\rangle$, $u=\langle D_{\xi_1}^{(i)}D_{\xi_2}^{(j)}|V_{ee}^{(i)(j)}|D_{\xi_2}^{(j)}D_{\xi_1}^{(i)}\rangle$, and 
\begin{equation}
\arraycolsep 0.3 ex
\begin{array}{rl}
\displaystyle H_{val} =\begin{pmatrix}
0 & 0 & \Delta_0\sqrt{2} & 0 \cr
0 & 0 & \Delta_0^*\sqrt{2} & 0 \cr
\Delta_0^*\sqrt{2} & \Delta_0 \sqrt{2} & \Delta_{ee} & 0 \cr
0 & 0 & 0 & -\Delta_{ee}
\end{pmatrix},
\end{array}
\end{equation}
with $\Delta_{ee}=\langle D_{z}^{(1)}D_{- z}^{(2)}|V_{ee}|D_{z}^{(2)}D_{- z}^{(1)}\rangle$. The eigenvalues of $H_{val}$ are $0$, $-\Delta_{ee}$ and  $\Delta_{\pm}=\Delta_{ee} \pm \sqrt{\Delta_{ee}^2 + 16|\Delta_0|^2}$, plotted in Fig.~\ref{fig:control} as functions of $\Delta_{ee}$. We note the crossover of eigenvalues in Fig.~\ref{fig:control}. At large $\Delta_0 \gg \Delta_{ee}$, the ground state is a spin singlet, yet when $\Delta_0$ is small enough, the ground state corresponds to the eigenvalue $-\Delta_{ee}$, and is a spin triplet. In such a case the ground state will move down in energy in a non-zero magnetic field.

To assess the possibility of a triplet ground state, we evaluate $\Delta_{ee}$. We expand $u_{z}({\bm r}) = \displaystyle\sum_{{\bm {K}}} c^z_{{\bm {K}}} e^{i{\bm {K}} \cdot{\bm r}}$, with ${\bm K}$ reciprocal lattice vectors, switch to center of mass and relative variables $\bm r= \bm r_1 -\bm r_2, \bm R= \bm r_1 +\bm r_2$, and employ cylindrical polar coordinates. The integration over ${\bm R}$ is trivial, while over ${\bm r} = (r_{\perp},\phi, z)$ we make the substitution $\frac 1 r=\frac 2 {\sqrt {\pi} }\int _{0}^{\infty}dte^{-r^2t^2}$, reducing the problem to
\begin{equation}
\Delta_{ee} = \frac{e^2\Sigma}{2\epsilon_0 \epsilon_r\sqrt{\pi}} \int_0^{\infty} \!\!\! du \, \cos(2k_0u)e^{\frac {u^2} {2a^2}}{\rm Erfc} \bigg(\frac u {a\sqrt{2}}\bigg)L(u),
\end{equation}
% We  $k_{\beta}=0$ ($\varsigma_{1z} +\varsigma_{2z} = 0$) because the final result of the integral should be a real number, while $k_{\beta}$ is a phase factor that may lead to a non-zero imaginary part. With $k_{\beta}=0$ we can further simplify our result.
with $\Sigma = \displaystyle \sum_{{\bm {K}}_1{\bm {K}}_2{\bm Q}_1{\bm Q}_2} c^{z*}_{{\bm {K}}_1} c^{-z*}_{{\bm {K}}_2} c^{-z}_{{\bm
{K}}_1 - {\bm Q}_1} c^z_{{\bm {K}}_2 - {\bm Q}_2}$, Erfc the complementary error function and
\begin{widetext}
\begin{equation}
\begin{array}{rl}
\displaystyle
L(u)=\frac{2^{\frac 1 2} \pi N^4z_0^4} {k_ba} e^{-k_bu}+
\frac {2\sqrt2\pi N^4z_0^2} {\delta_k^3a}e^{-k_bu}[e^{\delta_ku}\delta_k^2(u+z_0)^2-2e^{\delta_ku}\delta_k(u+z_0)+\delta_kz_0(2-\delta_kz_0)-2]+ \\ [3ex] \displaystyle
\frac {N^4\pi} {\sqrt2k_{Si}^5a}e^{-k_{Si}u}
[2k_{Si}^4z_0^2(u+z_0)^2+2k_{Si}^3z_0(u+z_0)(u+2z_0)+k_{Si}^2(u^2+6uz_0+6z_0^2)+3k_{Si}(u+2z_0)+3]
\end{array}
\end{equation}
\end{widetext}
Using $a=10$nm as the lowest realistic quantum dot radius, the effective $\epsilon_r=7.9$ for a sample Si/SiO$_2$ interface, the parameters of Ref.~\onlinecite{Culcer_ValleyQubit_PRL12} for the variational wave function and the coefficients $c^{\xi}_{\bm K}$ of Ref.~\onlinecite{Koiller_PRB04}, we obtain numerically $\Delta_{ee}=0.0352\mu eV$. The Coulomb interaction contribution to the VOC is negligible compared to the single-particle term $\Delta_0$, which is of the order of 0.05 meV \cite{Lim_SiQD_SpinFill_NT11}.

\subsection{Three-electron case}

We consider next ${N=3}$. The quantum dot can be in one of $C^4_3 = 4$ states. Using $|\uparrow\rangle$, $|\downarrow\rangle$ for up and down spinors respectively, one sample wave function is the Slater determinant
\begin{equation}
\arraycolsep 0.3 ex
\begin{array}{rl}
\displaystyle \phi^{3e}_1(z\uparrow,- z\uparrow,z\downarrow)= \frac 1 {\sqrt6} \begin{vmatrix}
D_{z\uparrow}^{(1)} & D_{- z\uparrow}^{(1)} & D_{z\downarrow}^{(1)} \cr
D_{z\uparrow}^{(2)} & D_{- z\uparrow}^{(2)} & D_{z\downarrow}^{(2)} \cr
D_{z\uparrow}^{(3)} & D_{- z\uparrow}^{(3)} & D_{z\downarrow}^{(3)} \cr
\end{vmatrix},
\end{array}
\end{equation}
and analogously $\phi^{3e}_2(z\uparrow,- z\downarrow,z\downarrow), \phi^{3e}_3(z\downarrow,- z\uparrow,- z\downarrow)$, and $\phi^{3e}_4(z\uparrow,-z\uparrow,- z \downarrow)$. The 3-electron Hamiltonian $H_{3e} = \sum_{i=1}^3T^{(i)}+V_D^{(i)} + \frac{1}{2}\sum_{i\neq j}^3 V_{ee}^{(ij)}$. 

Each basis state contains 6 terms, and each Hamiltonian matrix element has 36 terms. Spinor overlaps annihilate most of them, since the Coulomb interaction is spin-independent. We calculate $\langle \phi^{3e}_1|H_{3e} |\phi^{3e}_1 \rangle$ as an example. The six terms in $\phi^{3e}_1$ are $\phi_{11}=D_{z\uparrow}^{(1)}D_{- z\uparrow}^{(2)}D_{z\downarrow}^{(3)}$, $\phi_{12}=-D_{z\uparrow}^{(1)}D_{z\downarrow}^{(2)}D_{- z\uparrow}^{(3)}$, $\phi_{13}=D_{z\downarrow}^{(1)}D_{z\uparrow}^{(2)}D_{- z\uparrow}^{(3)}$, $\phi_{14}=-D_{z\downarrow}^{(1)}D_{- z\uparrow}^{(2)}D_{z\uparrow}^{(3)}$, $\phi_{15}=D_{- z\uparrow}^{(1)}D_{z\downarrow}^{(2)}D_{z\uparrow}^{(3)}$, and $\phi_{16}=-D_{- z\uparrow}^{(1)}D_{z\uparrow}^{(2)}D_{z\downarrow}^{(3)}$. We obtain
\begin{equation}
\arraycolsep 0.3ex
\begin{array}{rl}
\displaystyle \langle \phi_{11}|H_{3e}|\phi_{11}\rangle = & \displaystyle \Sigma_{i=1}^3\langle D_{z\uparrow}^{(1)}D_{- z\uparrow}^{(2)}D_{z\downarrow}^{(3)}|T^{(i)}|D_{z\uparrow}^{(1)}D_{- z\uparrow}^{(2)}D_{z\downarrow}^{(3)}\rangle \\ [1ex]

+ & \displaystyle 3\langle D_{z\uparrow}^{(1)}D_{- z- z\uparrow}^{(2)}D_{z\downarrow}^{(3)}|V_{ee}|D_{z\uparrow}^{(1)}D_{- z\uparrow}^{(2)}D_{z\downarrow}^{(3)}\rangle \\ [1ex]

= & \displaystyle 3\varepsilon_0+3u \\ [1ex]

\displaystyle \langle \phi_{11}|H_{3e}|\phi_{16}\rangle = & \displaystyle -\langle D_{z\uparrow}^{(1)}D_{- z\uparrow}^{(2)}D_{z\downarrow}^{(3)}|H_{3e}|D_{- z\uparrow}^{(1)}D_{z\uparrow}^{(2)}D_{z\downarrow}^{(3)}\rangle \\ [1ex]
= & \displaystyle -\Delta_{ee},
\end{array}
\end{equation}
and $\langle \phi_{11}|H_{3e}|\phi_{12}\rangle = \langle \phi_{11}|H_{3e}|\phi_{12}\rangle =\langle \phi_{11}|H_{3e}|\phi_{13}\rangle =\langle \phi_{11}|H_{3e}|\phi_{14}\rangle =\langle \phi_{11}|H_{3e}|\phi_{15}\rangle = 0$. Adding up these terms gives $\langle \phi^{3e}_1|H_{3e}|\phi^{3e}_1\rangle=3\varepsilon_0+3u-\Delta_{ee}$, and similarly $\langle \phi^{3e}_1|H_{3e}|\phi^{3e}_4\rangle = -\Delta_0$. The matrix elements $\langle \phi^{3e}_1|H_{3e}|\phi^{3e}_2\rangle$ and $\langle \phi^{3e}_1|H_{3e}|\phi^{3e}_3\rangle$ vanish because no terms in the bra and ket states share the same spin arrangement. Analogous arguments apply to all the remaining matrix elements, yielding
\begin{equation}
\arraycolsep 0.3 ex
\begin{array}{rl}
\displaystyle H_{3e} = & \displaystyle (3\varepsilon_0 + 3u-\Delta_{ee})\openone+\begin{pmatrix}
0 & 0 & 0 & -\Delta_0 \cr
0 & 0 & \Delta_0 & 0 \cr
0 & \Delta_0^{\ast} & 0 & 0 \cr
-\Delta_0^{\ast} & 0 & 0 & 0
\end{pmatrix}
\end{array}
\end{equation}

The finite matrix elements of $H_{3e}$ are $\langle\phi^{3e}_n|H_{3e}|\phi^{3e}_n\rangle=3\varepsilon_D+3u-\Delta_{ee}$ (for $1 <n < 4$) and $\langle\phi^{3e}_2|H_{3e}|\phi^{3e}_3\rangle=-\langle \phi^{3e}_1|H_{3e}|\phi^{3e}_4\rangle=\Delta_0$. The eigenvalues are $3\varepsilon_0 + 3u-\Delta_{ee} \pm\left| \Delta_0 \right|$. As a result of spinor overlaps, the Coulomb term $\Delta_{ee}$ does not appear in the off-diagonal matrix elements and does not contribute to the valley splitting, giving only an identical offset to all the energy eigenvalues. 

\begin{figure}[tbp]
\centering
\includegraphics[width=\columnwidth]{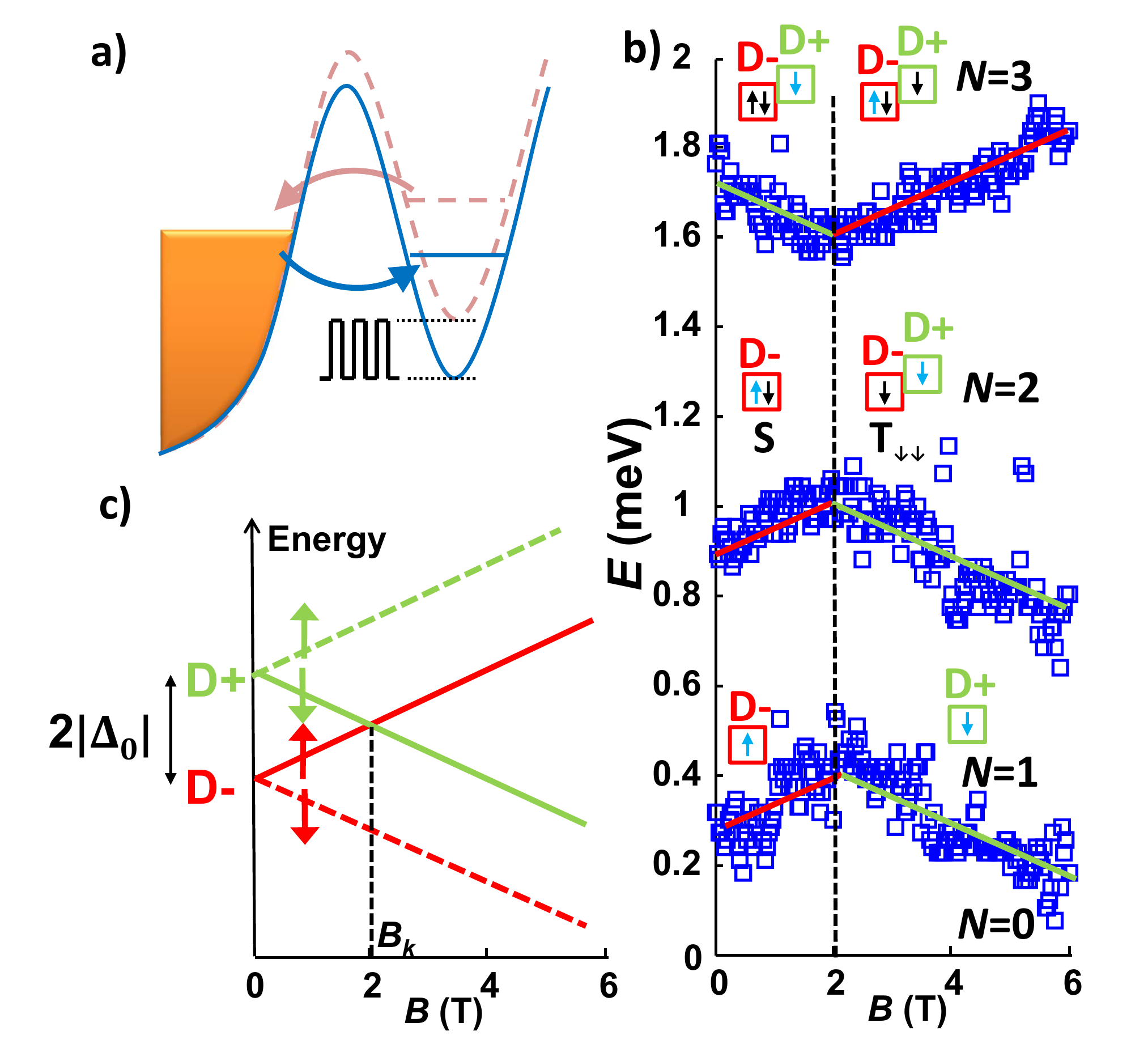}
\caption{(a) Energy diagrams of the dot when voltage pulses are superimposed to the DC bias. When a single-particle level is near the reservoir's Fermi level, an electron can be loaded into the dot during the upper phase of the pulse (solid lines) and unloaded during the lower phase (dashed lines). (b) Magnetic field dependence of the spin-valley states filling for $N \le 3$, where $D_\pm$ label single-electron valley eigenstates and $S$, $T_{\downarrow\downarrow}$ label spin states, consistent with Eq.~\ref{defs}. Data points show where the differential sensor's signal $\frac{dI_{SET}}{dE}$ is maximized. Solid lines are guides for the eye with slopes $+\frac{g\mu_B}{2}$ in red and $-\frac{g\mu_B}{2}$ in green. Boxes of different colors indicate different valley occupancies. Blue (black) arrows represent the spin state of the current (previous) electron addition(s). Sets of data for different charge transitions are arbitrarily shifted in energy for clarity. (c) Energy diagram of the 1-electron spin-valley states evolution in magnetic field. At the field value $B=B_k$ the two valley states cross, and Zeeman and valley-orbit splittings coincide. Dashed lines indicate states that are not probed in the experiments.}
\label{fig:expt}
\end{figure}

\subsection{Four-electron case}

For $N=4$, all four lowest-energy single-particle states are occupied, and only $C_4^4 = 1$ many-particle state exists. The Slater determinant wave function $\phi_{4e}$ for $N=4$ takes the form
\begin{equation}
\arraycolsep 0.3 ex
\begin{array}{rl}
\displaystyle \phi_{4e} = \frac 1 {\sqrt{24}} \begin{vmatrix}
D_{z\uparrow}^{(1)} & D_{- z\uparrow}^{(1)} & D_{z\downarrow}^{(1)} & D_{- z\uparrow}^{(1)}\cr
D_{z\uparrow}^{(2)} & D_{- z\uparrow}^{(2)} & D_{z\downarrow}^{(2)} & D_{- z\uparrow}^{(2)}\cr
D_{z\uparrow}^{(3)} & D_{- z\uparrow}^{(3)} & D_{z\downarrow}^{(3)} & D_{- z\uparrow}^{(2)}\cr
D_{z\uparrow}^{(4)} & D_{- z\uparrow}^{(4)} & D_{z\downarrow}^{(4)} & D_{- z\uparrow}^{(4)}\cr
\end{vmatrix}.
\end{array}
\end{equation}
The energy $\langle \phi_{4e}|H_{4e}|\phi_{4e}\rangle = 4\varepsilon_0+6u$ contains no intervalley terms, stemming from either the bare VOC or the Coulomb interaction, since interaction terms cancel between pairs of electrons with different spinors. 

\section{Experimental Spin-Valley Spectrum}
\label{sec:expt}

To experimentally probe the effect of Coulomb interactions on the VOC, we use quantum dots fabricated with metal-oxide-semiconductor (MOS) technology in nearly intrinsic Si \cite{Angus_NL07}. In these devices a three-layer Al-Al$_2$O$_3$-Al gate stack \cite{Lim_SingleElectron_APL09} allows one to electrostatically modulate the conduction band profile near the Si/SiO$_2$ interface. As a result, an electron accumulation layer is selectively formed in the semiconductor substrate and can function as either a quantum dot or a 2DEG reservoir, depending on the extent of planar confinement provided by the gates. The details of the fabrication process and typical bias configurations are reported elsewhere \cite{Lim_SiQD_SpinFill_NT11, Yang_OrbVal_PRB12}. A schematic diagram of the device used is shown in Fig.~\ref{fig:explain}(b). The system works as a single-lead quantum dot capacitively coupled to a single-electron-transistor (SET) acting as a charge sensor. Independent gate electrodes control the occupancy of the quantum dot, the tunnelling rate between the dot and the reservoir, the Fermi energy of the reservoir, and the bias point of the SET detector. The dot is reliably operated in the few-electron regime down to the last electron \cite{Yang_OrbVal_PRB12}.

The excitation spectrum of the quantum dot is measured using a pulsed voltage technique \cite{Elzerman_APL04}. A train of square voltage pulses at a frequency of few hundreds Hz is applied to the gate that directly affects the quantum dot potential (in addition to its DC voltage). This shifts the energy levels of the dot in time, inducing charge transitions whenever the dot's single-particle levels come into resonance with the lead's Fermi energy. An electron can be loaded from the reservoir into the dot during the upper phase of the pulse and unloaded during the lower phase (as shown in Fig.~\ref{fig:expt} (a)). The time-dependent gate voltage modulates the sensor current, $I_{SET}$, via capacitive effects, and a lock-in amplifier selects the spectral component which is uniquely associated to tunneling events (i.e. the one at the frequency of the pulse). This technique allows one to probe both ground and excited states for each $N$, as long as the relaxation rate is slower than the pulse frequency and the pulse magnitude spans the relevant energy separation. The measured spectrum reveals that the energy separation between the first excited orbital state and the ground state varies between 1-8meV according to dot's occupancy. A comprehensive description of the measurement set-up and parameters is found in Ref. \onlinecite{Yang_OrbVal_PRB12}. 

In Fig.~\ref{fig:expt}(b) measurements of the spin-valley states' evolution in a magnetic field ${\bm B} \parallel [110]$ (parallel to the Si/SiO$_2$ interface) are shown for $1 \le N \le 3$. Each data point indicates a maximum in the differential sensing current signal for varying DC gate voltage bias and a fixed pulse amplitude of 20 mV. The energy scale on the y-axis is obtained by converting the gate voltage into electrochemical potential using the appropriate lever arm conversion factor ($\approx$ 0.3 eV/V). The vertical shift of each trace is arbitrary and does not reflect the quantum dot charging energy. A strong sensing signal is due to charge transfers to/from the quantum dot. Hence, the observed magnetic field dependence maps the filling of spin-valley states for individual electrons. The slope of E(B) is given by \cite{Hada_PRB03}: $\frac{\delta E}{\delta B}=-g\mu_BB\Delta S_{tot}(N)$, where $g$ is the electron gyromagnetic ratio, $\mu_B$ is the Bohr magneton and $\Delta S_{tot}(N)$ is the change in total spin when the $N$-th electron is added to the dot. Hence, a slope of $+\frac{g\mu_B}{2}$ indicates a spin-up addition, while a $-\frac{g\mu_B}{2}$ slope is due to a spin-down electron. Solid lines shown in Fig.~\ref{fig:expt} (b) are guides for the eye with slopes $\pm \frac{g\mu_B}{2}$, using $g=2$ for bulk Si.

To understand the interplay between spin and valley degrees of freedom in a magnetic field, we analyse individual charge additions. The first transition ($N=0\rightarrow1$) reveals that for low magnetic fields the slope is positive up to $B=B_k\approx 2T$, where a change of sign occurs; hence, a spin-up (spin-down) electron is loaded into the dot for low (high) B-fields. For the first electron addition, one would expect to fill the spin-down lower valley eigenstate (the ground state) at all B-field values (see red dashed line in Fig.~\ref{fig:expt}(c)). However, the measurements indicate that the excited (ground) state of the lower $D_-$ (upper $D_+$) valley is probed for this transition for $B<B_k$ $(B>B_k)$. This occurs because the tunnelling rate between the dot and the lead in our measurement is faster than the relaxation rate, so that the largest contribution to the sensing signal comes from the loading/unloading of the excited states. Using Fig. ~\ref{fig:expt}(c) we can explain the presence of the kink (change of slope sign). For increasing magnetic fields the states $\ket{D_{\uparrow-}}$ and $\ket{D_{\downarrow+}}$ cross, making it energetically favourable for a spin-down electron to charge the dot. The position of the kink reveals the valley-orbit splitting, which coincides with the Zeeman splitting, as $g\mu_B B_k$ = 0.23 meV, yielding $|\Delta_0| = 0.115$ meV.

The trend in Fig.~\ref{fig:expt} (b) for the $N=1 \rightarrow 2$ transition is similar to that for $N=0\rightarrow1$. The ground state for $N=1$ is spin-down for all B. The spin-filling measurements reveal that for low B a spin-up electron is added to form a singlet which fully occupies the $-$ valley eigenstate. However, at larger B, a kink occurs at the same $B_k$ as for $N=1$. This suggests that for $B>B_k$ a spin-down electron is added to form the triplet $\ket{\phi^{T\downarrow\downarrow}_{mix}}$. The antisymmetry of the 2-electron wave-function requires one electron to occupy an excited state; here, this is the + valley eigenstate, which is significantly lower than the first excited orbital state \cite{Yang_OrbVal_PRB12}. The observed change of slope is due to a singlet-triplet crossing and, once again, the Zeeman and valley-orbit splittings coincide at $B=B_k$. 

For $N=2\rightarrow3$, we again observe a change of slope at $B=B_k$, but unlike the previous cases, the electron added at low (high) fields is spin-down (spin-up). To understand this, we note that for $B<B_k$ the 2-electron ground state is a spin singlet that fully occupies the lower valley eigenstate. The next available state would be a spin-down in the upper valley eigenstate. However, for $B>B_k$ the 2-electron ground state switches to $\ket{\phi^{T\downarrow\downarrow}_{mix}}$, which would allow one more electron in the lower valley eigenstate. This can only be spin-up because of Pauli exclusion. The change of slope at the same field value as for the previous transitions confirms that $|\Delta_0| = 0.115$ meV for $N=3$ as well. These measurements confirm that the VOC is independent of $N$ up to $N=3$, corroborating the prediction that $\Delta_{ee}$ is negligible. 

\section{Discussion}
\label{sec:disc}

The underlying physics of this problem is contained in the two-electron case. We have seen that the ground state for $N = 2$ is dependent on the magnitude of the quantity $\Delta_{ee}$, which can also be understood as representing \textit{intervalley exchange}. It is worth noting that $\Delta_{ee}$ is the \textit{only} exchange parameter appearing in this problem. The central message of this work is that there is a negligible gain in Coulomb energy for an electron to switch valley states. 

Fundamentally, $\Delta_{ee}$ is small because the Coulomb interaction varies on longer spatial scales than that set by the separation of the valleys in reciprocal space, which corresponds to $\approx 1\AA$ in real space. In other words, $\Delta_{ee}$ is suppressed by valley interference. Interaction effects are not expected to be observable even if $\Delta_0 = 0$, since $\Delta_{ee}$ corresponds to a few mK, while typical temperatures in a dilution refrigerator are $\approx $100mK. For any realistic Si two-dimensional electron gas (2DEG), the ground state of a multivalley two-electron single quantum dot is a spin singlet \cite{Lim_SiQD_SpinFill_NT11, Yang_SpinVal_NC13}.

For $N = 3$ and $N = 4$ we have no new matrix elements. We note, however, that the eigenvalues of the $N = 3$ case are doubly degenerate. Operating such a quantum dot as a qubit presents the same scalability issues as in Ref.~\onlinecite{Culcer_PRB10}. Briefly, although single qubit operations can be performed regardless of the state in which the system is initialized, two-qubit operations typically depend on exchange, which can be vastly different for different valley configurations.

\section{Summary}

In conclusion, we have demonstrated theoretically and experimentally that in few-electron quantum dots the VOC has no significant contribution from electron-electron interactions. We expect these findings to be generalizable to $N > 4$, since valley interference and spinor overlaps are equally relevant to higher orbital excited states. Conceptually, this proves that interactions are not sharp enough in real space to couple valleys, which have a large separation in ${\bm k}$-space. Together with previous research \cite{Saraiva_EMA_PRB11}, it shows that the VOC is fully determined by the properties of the interface: the size and shape of the interface potential step, electric field and roughness profile. 

These findings will aid experimentalists working on quantum computing in Si and other materials with valleys, such as C and Ge. The valley degree of freedom can be a significant impediment to single-spin and singlet-triplet qubits \cite{Culcer_PRB10}, and it is essential for their operation that the VOC be fully characterised, since valleys complicate the spin state spectrum on a fundamental level. The size of the VOC and its sensitivity to interactions and occupation number is also a critical ingredient of valley qubits \cite{Culcer_ValleyQubit_PRL12}, which may reduce sensitivity to noise.

\acknowledgements

We thank S.~Das Sarma, Xuedong Hu and R.~J.~Joynt for enlightening discussions. L.J., Z.P. and D.C. acknowledge support by the National Natural Science Foundation of China under grant number 91021019 and by LPS-NSA-CMTC.  C.H.Y., A.R. and A.S.D. acknowledge support from the Australian Research Council (project CE11E0096) and the U.S. Army Research Office (contract W911NF-13-1-0024). Experimental devices for this study were fabricated with support from the Australian National Fabrication Facility, UNSW.

% \bibliography{refs_Si}

\end{document}